\documentclass{article}

\usepackage{arxiv}

\usepackage[utf8]{inputenc} 
\usepackage[T1]{fontenc}    
\usepackage{hyperref}       
\usepackage{url}            
\usepackage{booktabs}       
\usepackage{amsfonts}       
\usepackage{nicefrac}       
\usepackage{microtype}      
\usepackage{lipsum}
\usepackage{listings}
\usepackage{color}
\usepackage[normalem]{ulem}
\usepackage{graphicx}

\lstdefinelanguage{scala}{
  morekeywords={abstract,case,catch,class,def,%
    do,else,extends,false,final,finally,%
    for,if,implicit,import,match,mixin,%
    new,null,object,override,package,%
    private,protected,requires,return,sealed,%
    super,this,throw,trait,true,try,%
    type,val,var,while,with,yield},
  otherkeywords={=>,<-,<\%,<:,>:,@},
  sensitive=true,
  morecomment=[l]{//},
  morecomment=[n]{/*}{*/},
  morestring=[b]",
  morestring=[b]',
  morestring=[b]"""
}
\definecolor{sh_comment}{rgb}{0.12, 0.38, 0.18 } 
\definecolor{sh_keyword}{rgb}{0.37, 0.08, 0.25} 
\definecolor{sh_string}{rgb}{0.06, 0.10, 0.98} 
\title{MaRe: a MapReduce-Oriented Framework for Processing Big Data with Application Containers}

\author{
  Marco Capuccini \\
  Department of Information Technology \\
  Uppsala University, Sweden \\
  \texttt{marco.capuccini@it.uu.se} \\
  \And
  Martin Dahlö \\
  Department of Pharmaceutical Biosciences \\
  Uppsala University, Sweden \\
  \And
  Salman Toor \\
  Department of Information Technology \\
  Uppsala University, Sweden \\
  \And
  Ola Spjuth \\
  Department of Pharmaceutical Biosciences \\
  Uppsala University, Sweden
}

\begin{document}
\maketitle

\begin{abstract}
\textbf{Background.}
Life science is increasingly driven by Big Data analytics, and the MapReduce programming model has been proven successful for data-intensive analyses. However, current MapReduce frameworks offer poor support for reusing existing processing tools in bioinformatics pipelines. Further, these frameworks do not have native support for application containers, which are becoming popular in scientific data processing.

\textbf{Results.}
Here we present MaRe, a programming model with an associated open-source implementation, which introduces support for application containers in MapReduce. MaRe is based on Apache Spark and Docker, the MapReduce framework and container engine that have collected the largest open source community, thus providing interoperability with the cutting-edge software ecosystem. We demonstrate MaRe on two data-intensive applications in life science, showing ease of use and scalability.

\textbf{Conclusions.}
MaRe enables scalable data-intensive processing in life science with MapReduce and application containers. When compared with current best practices, that involve the use of workflow systems, MaRe has the advantage of providing data locality, ingestion from heterogeneous storage systems and interactive processing. MaRe is generally-applicable and available as open source software.
\end{abstract}

\keywords{MapReduce \and application containers \and Big Data \and Apache Spark \and workflows.}

\section{Findings}
\subsection{Background and purpose}
Life science is increasingly driven by Big Data analytics. From genomics, proteomics and metabolomics to bioimaging and drug discovery, scientists need to analyze larger and larger amounts of data \cite{stephens2015big,foster2017intersection,peters2018phenomenal,peng2008bioimage,brown2018big}. This means that datasets can no longer be stored and processed in a researcher's workstation, but they instead need to be handled on distributed systems, at organization level. For instance, the European Bioinformatics Institute, in Hinxton (United Kingdom), offers a total storage capacity of over 160 petabytes for biologically-significant data \cite{cook2018european}. Such amounts of data poses major challenges for scientific analyses.
First, there is a need to efficiently scale existing processing tools over massive datasets. In fact, bioinformatics software that was originally developed with the simplistic view of small-scale data, will not scale on distributed computing platforms out of the box. The process of adapting such tools may introduce disruptive changes to the existing codebase, and it is generally unsustainable for most organizations. 
Secondly, the complexity in programming distributed systems may be hard to cope with for most researchers, who instead need to focus on the biological problem at hand. In addition, as life science is exploratory, scientists increasingly demand being able to run interactive analyses rather than submitting jobs to batch systems. 
Thirdly, when handling Big Data in distributed systems, data locality is a major concern. Indeed, if once data could be shuffled with little regard, with massive datasets it is not only inefficient \cite{tan2012delay}, but also prohibitively expensive in terms of power consumption, estimated to be in the order of several hundred thousand dollars per year \cite{still2015gearing}. For geographically dispersed datasets, locality-awareness becomes even more challenging, as computing resources need to be dynamically acquired close to the data \cite{convolbo2018geodis}. Cloud computing solves this problem by enabling the allocation of virtual infrastructure on demand \cite{fox2009above}. However, heterogeneity in storage systems for cloud providers \cite{mansouri2018data}, makes it hard to abstract data ingestion from many different sources.
Finally, as bioinformatics software is characterized by complex software dependencies, deploying and managing a vast collection of tools in a large distributed system also represents a major challenge \cite{williams2016growing}.

Current bionformatics best practices make use of workflow systems, to orchestrate analyses over distributed computing platforms \cite{leipzig2017review}. Workflow systems provide high-level Application Programming Interfaces (APIs) that allow for defining an execution graph of existing processing tools. At run time, the execution graph is used to pipeline the analysis on distributed cloud or High-Performance Computing (HPC) resources. Hence, the parallelization of the analysis is transparently carried out, by executing non-dependent tasks at the same time. 
Cutting-edge workflow systems, such as Luigi \cite{lampa2016towards}, NextFlow \cite{di2017nextflow}, Galaxy \cite{moreno2018galaxy} and Pachyderm \cite{novella2018container} allow for running processing tools as application containers. This light-weight packaging technology allows for encapsulating complete software environments, so that distributed systems can run the processing tools with no need of additional dependencies, in an isolated manner \cite{oci2016}. Hence, container-enabled workflow systems provide a fairly easy way to define distributed analyses comprising existing bioinformatics tools, and eliminating the need for managing complex software delivery process and dependency management.
Nevertheless, workflow-oriented processing falls short when it comes to Big Data analyses. To the best of the authors knowledge, all of these systems utilize a decoupled shared storage system, for synchronization and intermediate results storage. When dealing with large datasets, this translates to a massive and unnecessary communication in the underlying infrastructure. In addition, workflow systems usually support a limited amount of storage backends, not seldom only POSIX file systems, making it hard to ingest data from heterogeneous cloud resources. Finally, due to their batch-oriented nature, it is also intrinsically hard to enable interactive, exploratory analyses using workflow-oriented frameworks.

Google's MapReduce programming model and its associated implementation pioneered uncomplicated Big Data analytics on distributed computing platforms \cite{dean2008mapreduce}. When using MapReduce, the analysis is defined in a high-level programming language that hides challenging parallel programming details including fault tolerance, data distribution and locality-aware scheduling. Open-source implementations of MapReduce are well established in industrial and scientific applications~\cite{bhandarkar2010mapreduce, gunarathne2010mapreduce}, and numerous success stories in life science have been reported~\cite{mohammed2014applications,guo2018bioinformatics,schonherr2012cloudgene}.

Apache Spark has emerged as the project that collected the largest community, in the open-source MapReduce ecosystems \cite{zaharia2016apache}. In addition to the MapReduce implementation, Apache Spark also provides increasingly important features, such as in-memory, interactive and stream processing. Furthermore, due to broad collaborations in the open-source community, Apache Spark supports all of the major storage systems, enabling data ingestion from heterogeneous cloud resources.
These characteristics are particularly appealing for the case of Big Data in life science. Nevertheless, Apache Spark, and other similar frameworks, offer poor support for composing analyses out of existing processing tools. This is usually limited to calling external programs, which can only access data sequentially, without support for application containers \cite{ding2011more}. In fact, the main way of implementing analytics in MapReduce-oriented environments is to code each transformation using one of the available APIs. This way of implementing analyses contrasts with current best practices in bioinformatics, that promote the usage of existing tools as application containers with the goal of improving delivery, interoperability and reproducibility of scientific pipelines \cite{di2017nextflow}.

Here we introduce MaRe: an open-source programming library that extends Apache Spark, introducing comprehensive support for external tools and application containers in MapReduce. Similarly to container-enabled workflow systems, MaRe allows to define analyses in a high-level language, in which data transformations are performed by application containers. In addition, MaRe provides seamless management of data locality as well as full interoperability, with the Apache Spark ecosystem. This last point allows MaRe analyses to ingest data from heterogeneous cloud storage systems, and also provides support interactive processing. Finally by supporting Docker, the de facto standard container engine \cite{shimel2016docker}, MaRe is compatible with numerous existing container images.

In summary, the key contribution of the presented work are:

\begin{itemize}
    \item We introduce a MapReduce-oriented programming model for container-based data processing.
    \item We provide MaRe: an open-source implementation of the introduced programming model.
    \item We benchmark MaRe on two data-intensive applications in life science, showing ease of use and scalability, also in relation to three large-scale storage systems.
\end{itemize}

\subsection{MaRe}

\subsubsection{Programming Model}
We introduce the MaRe programming model using a simple, yet interesting, example in genomics. A DNA sequence can be represented as a text file written in a language of 4 characters: A,T,G,C. The GC content in a DNA sequence has interesting biological implications; for instance there is evidence that GC-rich genes are expressed more efficiently than GC-poor genes \cite{kudla2006high}. Hence, within a large DNA sequence it can be interesting to count G and C occurrences. Given an Ubuntu Docker image \cite{ubuntuDocker}, the task can easily be implemented in MaRe using POSIX tools. Listing \ref{lst:gc} shows such implementation.

\begin{lstlisting}[language=scala,caption=GC count in MaRe, label=lst:gc]
val gcCount = new MaRe(genomeRDD).map(
  inputMountPoint = TextFile("/dna"),
  outputMountPoint = TextFile("/count"),
  imageName = "ubuntu",
  command = """
    grep -o '[GC]' /dna | wc -l > /count
  """
).reduce(
  inputMountPoint = TextFile("/counts"),
  outputMountPoint = TextFile("/sum"),
  imageName = "ubuntu",
  command = """
    awk '{s+=$1} END {print s}' /counts > /sum
  """
)
\end{lstlisting}

Being based on Apache Spark, MaRe has a similar programming model. The control flow of the analysis is coded in Scala \cite{odersky2004overview}, by the program in listing \ref{lst:gc}. Such program is called \textit{driver} in the Apache Spark terminology. The driver program can be packaged and submitted to a cluster (in batch mode), or executed interactively using a notebook environment such as Jupyter \cite{kluyver2016jupyter} or Apache Zeppelin \cite{cheng2018building}. Listing \ref{lst:gc} starts by instantiating a \texttt{MaRe} object, which takes a Resilient Distributed Dataset (RDD) \cite{zaharia2012resilient}, containing the input genome file in text format. Such RDD can be easily loaded using the Apache Spark API from any of the supported storage backends. The \texttt{map} primitive (line 1 to 8) applies a command from the Docker image to each partition of the RDD. In our example we specify the Ubuntu image on line 4, and we use a command that combines \texttt{grep} and \texttt{wc} to filter and count GC occurrences (on line 6). The partitions are mounted in the Docker containers in the configured input mount point (\texttt{"/dna"} at line 2), and the command results are loaded back to MaRe from the configured output mount point (\texttt{"/count"} on line 3). In the example we use \textit{TextFile} mount points as the input data is in text format. By default, MaRe considers each line in a text file as a separate record, but custom record separators can also be configured using the \textit{TextFile} constructor.

At this point it is important to mention that MaRe can also handle binary files. For such data formats, the driver program should specify mount points of type \texttt{BinaryFiles}. In this case, each RDD record is considered as a distinct binary file, thus the specified mount point results in a directory containing multiple files (as opposed to \texttt{TextFile} that mounts the records in a single file). We provide an example of the \texttt{BinaryFiles} mount point in the evaluation section.

Coming back to listing \ref{lst:gc}, after applying the \texttt{map} primitive, each RDD partition is transformed into a distinct GC count. The \texttt{reduce} primitive (line 8 to 15), aggregates the counts in each partition to a cumulative sum. Again, we use mount points of type \texttt{TextFile}, to mount the intermediate counts in the containers (\texttt{"/counts"} on line 9) and to read back the cumulative sum (\texttt{"/sum"} on line 10). The sum is computed using the \texttt{awk} command from the Ubuntu image (lines 11 to 14). Finally, the result is returned to the \texttt{gcCount} variable at line 1.

From the GC example, the reader may have noticed that our programming model is strongly inspired by MapReduce. In addition, Apache Spark users may have noticed that the GC count problem can easily be solved in pure Spark code. Indeed, the aim of the example is just to provide an easy introduction to MaRe, and two real-world applications are available in the evaluation section. 

Apart from \texttt{map} and \texttt{reduce}, MaRe provides an additional primitive. For real-world applications, we noticed that it is often needed to group dataset records according to a specific logic before applying \texttt{map} or \texttt{reduce}. For this reason, MaRe also provides a \texttt{repartitonBy} primitive, which repartitions the RDD records according to a configurable grouping rule. More specifically, the \texttt{repartitonBy} primitive takes into account a user-provided \texttt{keyBy} function, which is used to compute a key for each record in the dataset. Then, the repartitioning is performed accordingly so that records with same key end up in the same partition. An example of \texttt{repartitonBy} is available in the evaluation section.

\subsubsection{Implementation}
MaRe comes as a thin layer on top of the RDD API \cite{zaharia2012resilient}, and it relies on Apache Spark to provide important features such as data locality, data ingestion, interactive processing, and fault tolerance. The implementation effort consists of: (i) leveraging the RDD API to implement the MaRe primitives and (ii) handling data between containers and RDD structures. 

\paragraph{Primitives}
Each instance of a \texttt{MaRe} object retains an underlying RDD, which represents an abstraction of a dataset that is partitioned across Apache Spark workers. The \texttt{map}, \texttt{reduce} and \texttt{repartitionBy} primitives utilize the underlying RDD API to operate such dataset.

\begin{figure*}
    \centering
    \includegraphics[width=0.5\textwidth]{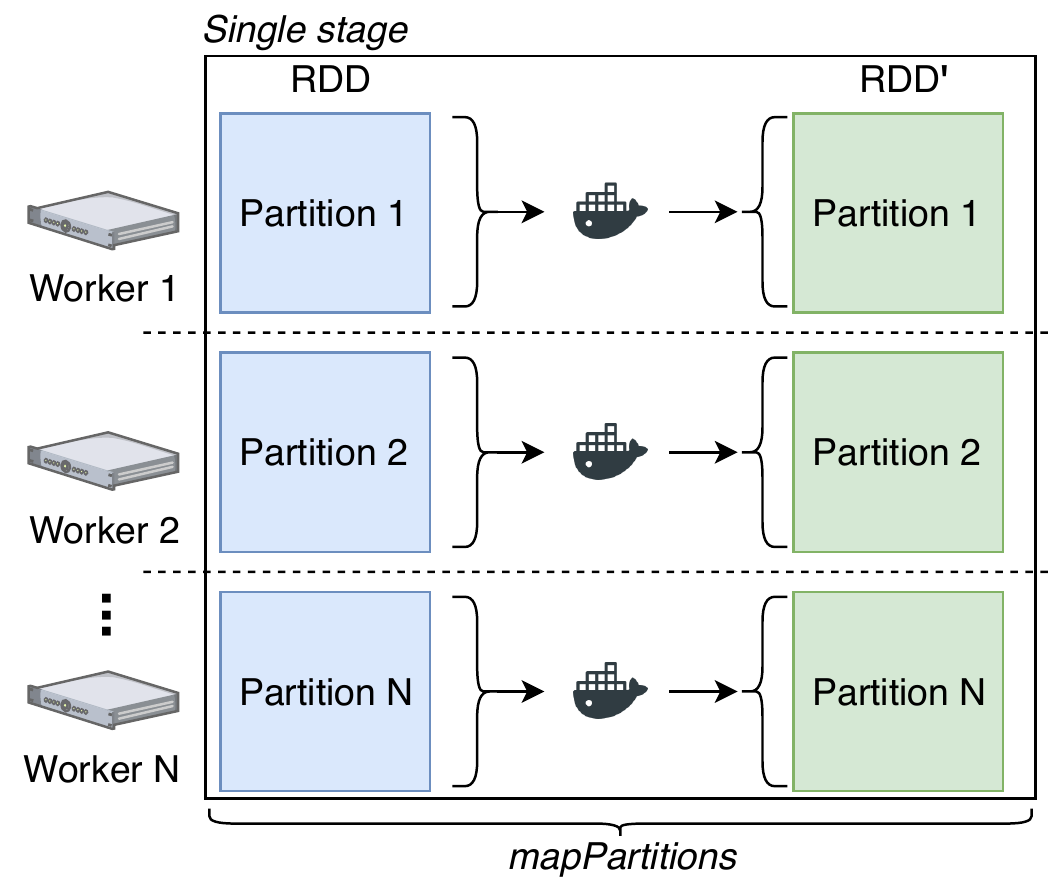}
    \caption{Execution diagram for the \texttt{map} primitive. The primitive takes an RDD that is partitioned over N nodes, it transforms each partition using a Docker container and it returns a new RDD'. The logic is implemented using \textit{mapPartitions} from the RDD API. Since \textit{mapPartitions} generates a single stage, data is not shuffled between nodes.}
    \label{fig:MaRe_map}
\end{figure*}

Figure \ref{fig:MaRe_map} shows the execution diagram for the \texttt{map} primitive. For simplicity, in figure \ref{fig:MaRe_map} we show a single partition per worker, but in reality workers may retain multiple partitions. This primitive takes an input RDD that is partitioned over N nodes, and it transforms each partition using a Docker container command - thus returning a new RDD'. This logic is implemented using \textit{mapPartitions} from the RDD API. When calling \textit{mapPartitions}, MaRe specifies a lambda expression that: (i) makes the data available in the input mount point, (ii) runs the Docker container and (iii) retrieves the results from the output mount point. When using \textit{mapPartitions}, Apache Spark generates a single stage, thus no data shuffle is performed. 

\begin{figure*}
    \centering
    \includegraphics[width=\textwidth]{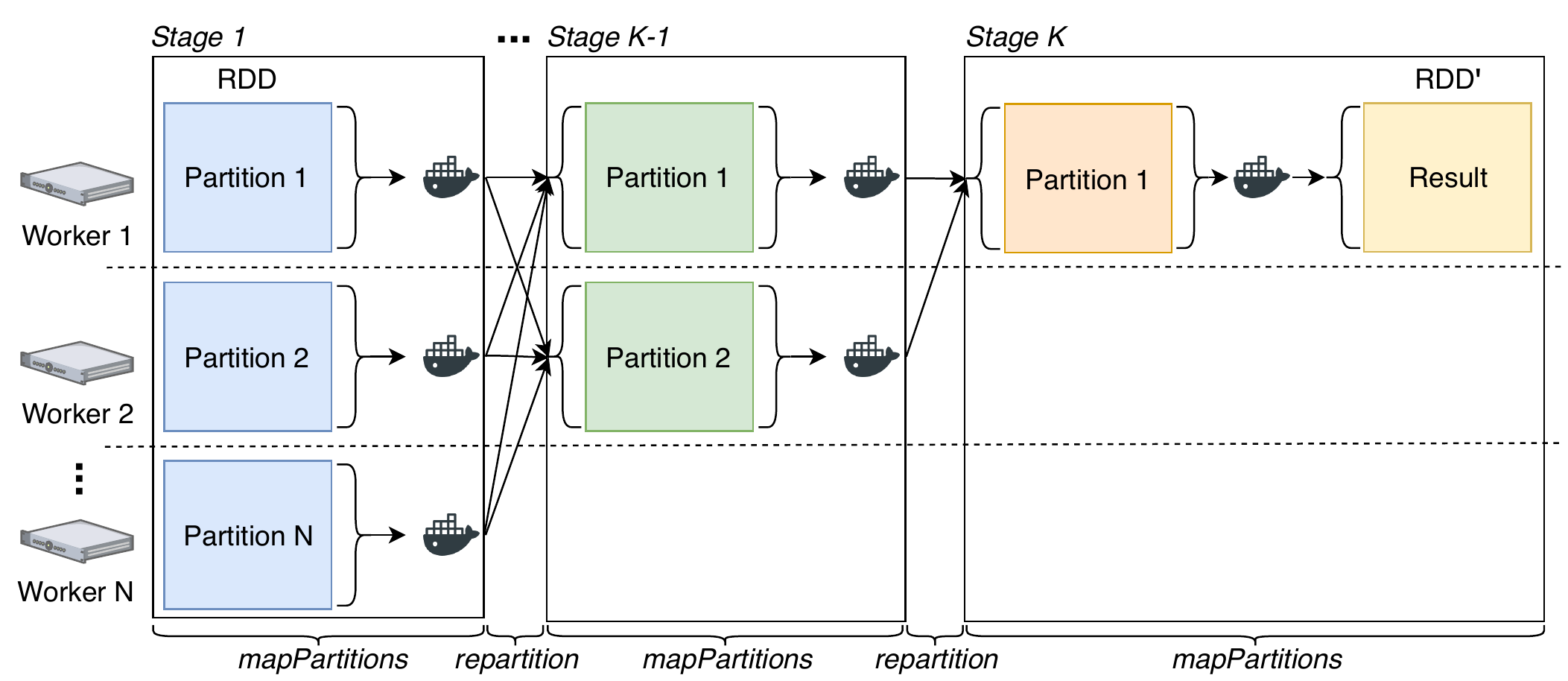}
    \caption{Execution diagram for the \texttt{reduce} primitive. The primitive takes an input RDD, partitioned over N nodes, and it iteratively aggregates records using a Docker container, reducing the number of partition until an RDD’, containing a single result partition, is returned. The logic is implemented using \textit{mapPartitions} and \textit{repartition} from the RDD API, to aggregate records in partitions and to decrease the number of partitions respectively. Since \texttt{repartition} is called in each of the K iterations, K stages are generated, giving place to K data shuffles.}
    \label{fig:MaRe_reduce}
\end{figure*}

Figure \ref{fig:MaRe_reduce} shows the execution diagram for the \texttt{reduce} primitive. This primitive takes an input RDD, partitioned over N nodes, and it iteratively aggregates records, reducing the number of partition until an RDD', containing a single result partition, is returned. Again, the input RDD may retain multiple partitions per node. However, as opposed to the \texttt{map} primitive, RDD' always contains a single partition when it is returned. Given a user-configured depth K, the records in the RDD are aggregated using a tree-like algorithm. In each of the K levels in the tree, the records within each partitions are first aggregated using a Docker container command. Like the \texttt{map} primitive, this first transformation is implemented using \textit{mapPartitions}, from the RDD API. Then, the number of partitions is decreased using \textit{repartition} from the RDD API. This process is repeated K times until one single partition is left. At this point the records within the remaining partition are aggregated again using \textit{mapPartitions} (from the RDD API), and RDD' is returned. A new stage is generated each time \textit{repartition} is used. Hence, \texttt{reduce} leads to K data shuffles. For this reason, when aggregating records, the user-provided command should always reduce the size of the partition. In addition, for results consistency, the command should perform an associative and commutative operation. By default MaRe sets K to 2, however the user may chose a higher tree depth when it is not possible to sufficiently reduce the dataset size in one go.

Finally, the \texttt{repartitionBy} primitive is implemented by using \textit{keyBy}, and then \textit{repartition} from the RDD API. MaRe uses the user-provided grouping rule with \textit{keyBy}, to compute a key for each RDD record, and then it applies \texttt{repartition} in counjuction with \textit{HashPartitioner} \cite{hashpartitioner}, which makes sure that records with same key end up in the same partition.

\paragraph{Data Handling}
One of the advantages of Apache Spark over other MapReduce-like systems is the ability of retaining data in memory. Hence, for better performance it is preferable to keep RDD records in memory when mounting them in the containers. To achieve this, there are a few options available: (i) Unix pipes \cite{peek1998unix}, (ii) memory-mapped files \cite{tevanian1987unix} and (iii) \textit{tmpfs} \cite{snyder1990tmpfs}. Solution (i) and (ii) are the most performant as they do not need to materialize the data when passing it to the containers. However, (i) allows to see records only once in a stream-like manner, while (ii) requires the container-wrapped tools to be able to read from a memory-mapped file. Therefore, to support any wrapped-tool, we decided to start by implementing solution (iii). This means that MaRe uses an in-memory \textit{tmpfs} file system as temporary file space for the input and output mount points. The solution allows to provide a standard POSIX mount point to the containers, while still retaining reasonable performance \cite{snyder1990tmpfs}. However, MaRe also provides users with the option of selecting any other disk-based file system for the temporary mount points. Even if this is not the best solution from a performance perspective, it can still be useful for particularly large partitions that need to be processed at once.

\subsection{Evaluation}
We evaluate MaRe on two data-intensive applications in life science. More specifically we evaluate: (i) how the analyses can be implemented in MaRe and (ii) how the analyses scale over multiple nodes.

The scalability experiments were carried out on cPouta: an OpenStack-based cloud service operated by the Information Technology Center for Science (CSC) in Finland \cite{cpouta}. On top of cPouta, we ran a stand-alone Apache Spark cluster composed of 1 master and 16 worker nodes. Each node provided 8 virtual Central Processing Units (vCPUs) and 32GB of memory, thus resulting in a total of 128 vCPUs and 512GB of memory. The driver programs were run interactively using an Apache Zeppelin environment \cite{cheng2018building}, and the notebooks were made available to sustain reproducibility~\cite{benchmarks}. In addition, we also made available a deployment automation that enables to replicate our setup on cPouta, as well as any other OpenStack-based cloud provider \cite{tf_spark}.

Since MaRe is conceived for data-intensive applications, the scalability is primarily evaluated in terms of Weak Scaling Efficiency (WSE). This performance metric shows how the system scale when increasing data and parallelism. To compute the WSEs we first ran the analyses on the full evaluation datasets using 16 worker nodes. Then, we ran again on 1/2, 1/4, 1/8 and 1/16 of the datasets, using 8, 4, 2 and 1 nodes respectively. The WSE is then computed as the time for processing the 1/16 of the data on 16 nodes, divided by the time for processing 1/N of the data using 16/N nodes (for N=1,2,4,8,16). The ideal case, when doubling the number of nodes, is to be able to process twice as much data in the same amount of time. Hence, a higher WSE indicates better performance.

We demonstrate data ingestion from three large-scale storage systems: Hadoop Distributed File System (HDFS) \cite{shvachko2010hadoop}, Swift \cite{swift} and Amanzon S3 \cite{s3}. In our settings HDFS was co-located with the Apache Spark cluster. This means that the HDFS daemons ran in the worker nodes, allowing for near-zero network communication. Swift is provided as a service by cPouta, thus being decoupled from our worker nodes. However, by setting up the cluster on cPouta, we ran the analyses close to Swift (thus enabling fast ingestion). Finally, S3 is provided by Amazon, hence in this case the analysis accessed data from a remote location. Even if this is not optimal from a performance perspective, it can sometimes be unfeasible to store large datasets locally and here we aim to show the tradeoff for this setting. 

\subsubsection{Virtual Screening}
\label{sec:vs}
Virtual Screening (VS) is a computer-based method to identify potential drug candidates, by evaluating the binding affinity of virtual compounds against a biological target protein \cite{cheng2012structure}. Given a 3D target structure, a molecular docking software is run against a large library of known molecular representations. For each compound in the virtual molecular library the docking software produces a pose, representing the orientation of the molecule in the target structure, and a binding affinity score. The poses with the highest affinity scores can be considered as potential drug leads for the target protein.

VS is data-intensive as molecular libraries usually contain millions of compounds. A simple, yet effective, approach to scale VS consists of: (i) distributing the molecular library over several nodes, (ii) running the docking software in parallel and (iii) aggregating the top-scoring poses. Listing \ref{lst:vs} shows how this logic can be implemented in MaRe, using FRED \cite{mcgann2011fred} as molecular docking software, and sdsorter \cite{sdsorter} to filter the top-scoring poses.

\begin{lstlisting}[language=scala,caption=Virtual Screening in MaRe, label=lst:vs]
val topPosesRDD = new MaRe(libraryRDD).map(
  inputMountPoint = TextFile("/in.sdf", "\n$$$$\n"),
  outputMountPoint = TextFile("/out.sdf", "\n$$$$\n"),
  imageName = "mcapuccini/oe:latest",
  command = """
    fred -receptor /var/openeye/hiv1_protease.oeb \
      -hitlist_size 0 \
      -conftest none \
      -dbase /in.sdf \
      -docked_molecule_file /out.sdf
  """
).reduce(
  inputMountPoint = TextFile("/in.sdf", "\n$$$$\n"),
  outputMountPoint = TextFile("/out.sdf", "\n$$$$\n"),
  imageName = "mcapuccini/sdsorter:latest",
  command = """
    sdsorter -reversesort="FRED Chemgauss4 score" \
      -keep-tag="FRED Chemgauss4 score" \
      -nbest=30 \
      /in.sdf /out.sdf
  """
)
\end{lstlisting}

In listing \ref{lst:vs}, we initialize MaRe by passing it a molecular library that was previously loaded as an RDD (\texttt{libraryRDD} on line 1). We implement the parallel molecular docking using the \texttt{map} primitive. On line 2 and 3, we set input and output mount points as text files, and assuming the library to be in Structure-Data File (SDF) format \cite{dalby1992description} we use the custom record separator: "\texttt{\textbackslash n\$\$\$\$\textbackslash n}". On line 4, we specify a Docker image containing FRED. The image is not publicly available as it also contains our FRED license, but the license can be obtained free of charge for research purposes and we provide a \textit{Dockerfile} \cite{benchmarks} to build the image. On line 5, we specify the FRED command. We use a HIV-1 protease receptor \cite{backbro1997unexpected} as target (which is wrapped in the Docker image), and we set: (i) \texttt{-hitlist\_size 0} to not filter the poses in this stage, (ii) \texttt{-conftest none} to consider the input molecules as single conformations, (iii) \texttt{-dbase /in.sdf} to read the input molecules from the input mount point and (iv) \texttt{-docked\_molecule\_file /out.sdf} to write the poses to the output mount point.

The map phase produces a pose for each molecule in \texttt{libraryRDD}. On line 12, we use the \texttt{reduce} primitive to filter the top 30 poses. On line 13 and 14, we set the input and output mount points as we do for the \texttt{map} primitive. On line 15, we specify a publicly available Docker image containing sdsorter. On line 16, we specify the sdsorter command, and we set: (i) \texttt{-reversesort="FRED Chemgauss4 score"} to sort the poses from highest to lowest FRED score, (ii) \texttt{-keep-tag="FRED Chemgauss4 score} to keep the score in the results, (iii) \texttt{-nbest=30} to output the top 30 poses and (iv) \texttt{/in.sdf /out.sdf} to read and write from the input mount point and to the output mount point respectively. Please notice that this command performs an associative and commutative operation, thus ensuring correctness in the reduce phase. Finally, the results are returned to \texttt{topPosesRDD}, on line 1.

\begin{figure*}
    \centering
    \includegraphics[width=0.5\textwidth]{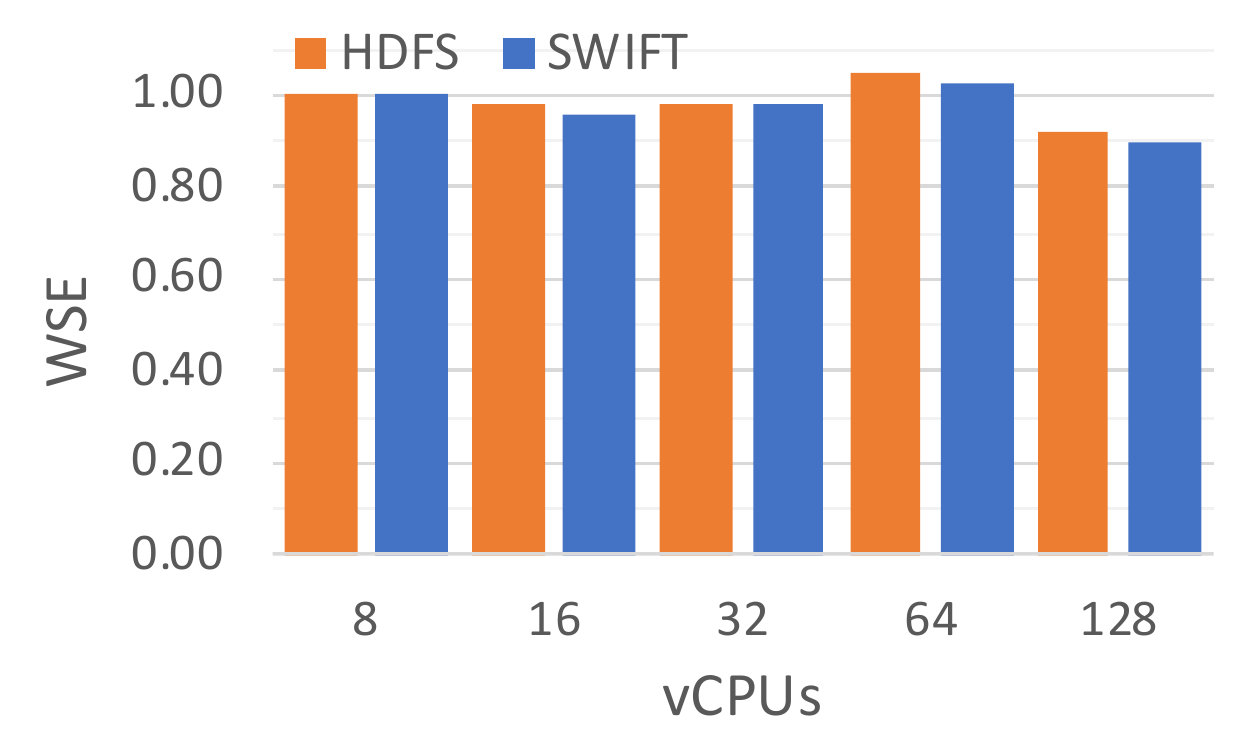}
    \caption{WSE for the VS application implemented in MaRe (listing \ref{lst:vs}). The results are produced by using SureChEMBL as input, and we show the WSE for two storage backends: HDFS and Swift. Please notice that the \textit{vCPUs} axis is in base two logarithmic scale.}
    \label{fig:vs_wse}
\end{figure*}

We benchmarked the analysis coded in listing \ref{lst:vs} against the SureChEMBL library \cite{papadatos2015surechembl} retrieved from the ZINC database \cite{irwin2012zinc}, containing $\sim$2.2M molecules. The full benchmark runs in $\sim$3 hours when using 128 vCPUs. Figure \ref{fig:vs_wse} shows the WSE for the full analyisis, when using HDFS and Swift. The results in figure \ref{fig:vs_wse} indicate very good scalability, with WSE close and even exceeding 1, up to 64 vCPUs. For 128 vCPUs the WSE levels off slightly at $\sim$0.9, still indicating good scalability. Finally, to ensure the correctness of the parallelization, we ran sdsorter and FRED on a single core against 1K molecules that we randomly sampled from SureChEMBL, and we compared the results with those produced by the code in listing \ref{lst:vs}.

\subsubsection{Single Nucleotide Polymorphism Calling}
\label{sec:snp}
A Single Nucleotide Polymorphism (SNP) is a position in a DNA sequence where a single nucleotide (or base pair) is different when compared to another DNA sequence \cite{karki2015defining}. When considering multiple samples, DNA sequences are usually compared individually to a reference genome: an agreed-upon sequence that is considered to represent an organisms genome. Once each DNA sequence has had its SNPs detected, or \textit{called}, the differences between the samples can be compared.

SNPs are frequently occurring. In fact, in humans roughly every 850th base pair is a SNP~\cite{10002015global}. Calling SNPs has several use cases. For instance, SNPs can be used as high-resolution markers when comparing genomic regions between samples \cite{Collins1999}, as well as indicators of diseases in an individual \cite{Kruglyak1999}. Modern high-throughput sequencing methods for reading DNA often make use of a technique called \textit{massively parallel sequencing}, to read sequences longer than $\sim$200 base pairs, with a sufficiently small error rate.  This is done by cleaving multiple copies of the source DNA into random fragments (called \textit{reads}) that are small enough to be accurately read, and then by aligning them to a reference genome. The overlapping fragments together form the sequence of the source DNA.

In order to accurately sequence 3 billion bases from a single human individual, 30-fold more reads data needs to be sequenced \cite{stephens2015big}. This makes SNP calling data-intensive, thus requiring parallelization. A simple MapReduce-oriented approach consists of: (i) distributing the reads across several nodes, (ii) aligning the reads to a reference genome in parallel and (iii) calling the SNPs with respect to the reference genome. The last step requires all the reads from a chromosome to be included in the SNP calling, thus the maximum allowed parallelism is equal to the total number of chromosomes. Listing \ref{lst:snp} shows how the described parallelization can be implemented in MaRe, using BWA for the alignment \cite{li2009fast} and GATK \cite{mckenna2010genome} for the SNP calling. As opposite the VS example, BWA and GATK provide a multithreaded implementation of the algorithms. Therefore, in listing \ref{lst:snp}, we leverage such implementation for single-node parallelization.



\begin{lstlisting}[language=scala,caption=SNP Calling in MaRe, label=lst:snp]
val snpRDD = new MaRe(readsRDD).map(
  inputMountPoint = TextFile("/in.fastq"),
  outputMountPoint = TextFile("/out.sam"),
  imageName = "mcapuccini/alignment:latest",
  command = """
    bwa mem -t 8 \
      -p /ref/human_g1k_v37.fasta \
      /in.fastq \
      | samtools view > /out.sam
  """
).repartitionBy(
    keyBy = (sam: String) => parseChromosomeId(sam),
    numPartitions = numberOfNodes
).map(
  inputMountPoint = TextFile("/in.sam"),
  outputMountPoint = BinaryFiles("/out"),
    imageName = "mcapuccini/alignment:latest",
    command = """
      cat /ref/human_g1k_v37.dict /in.sam \
        > /in.hdr.sam
      gatk AddOrReplaceReadGroups \
        --INPUT=/in.hdr.sam \
        --OUTPUT=/in.hdr.sort.rg.bam \
        --SORT_ORDER=coordinate \
        [ ... header options ... ]
      gatk BuildBamIndex \
        --INPUT=/in.hdr.sort.rg.bam
      gatk HaplotypeCallerSpark \
        -R /ref/human_g1k_v37.fasta \ 
        -I /in.hdr.sort.rg.bam \
        -O /out/${RANDOM}.g.vcf
      gzip /out/*
  """
).reduce(
    inputMountPoint = BinaryFiles("/in"),
    outputMountPoint = BinaryFiles("/out"),
    imageName = "opengenomics/vcftools-tools:latest",
    command = """
      vcf-concat /in/*.vcf.gz \
        | gzip -c > /out/merged.${RANDOM}.g.vcf.gz
    """
)
\end{lstlisting}

In listing \ref{lst:snp}, MaRe is initialized by passing an RDD containing the reads for a human individual in interleaved FASTQ format\cite{Cock2010} (\texttt{readsRDD} on line 1). We implement the parallel reads alignment using the \texttt{map} primitive. From line 2 to 4, we set the mount points as text files, and we specify a publicly available Docker image containing the necessary software tools. On line 5 we specify the BWA command and we set: (i) \texttt{-t 8} to utilize 8 threads, (ii) \texttt{-p /ref/human\_g1k\_v37.fasta} to specify the reference genome location (in the container) and (iii) the input mount point \texttt{/in.fastq}. In addition, on line 9 we pipe the results to another software, called \texttt{samtools} \cite{Li2009}, to convert them from the binary BAM format \cite{Li2009} to the text SAM format \cite{Li2009}. Converting the results to text format makes it easier to parse the chromosome location in the next step.

When calling SNPs, GATK needs to read all of the aligned reads for a certain DNA region. Using chromosomes to define the regions makes sure that no reads will span a region break point - a problem that would need to be handled if chromosomes were to be split in smaller regions. To achieve this we need to: (i) perform a chromosome-wise repartition of the dataset and (ii) allow MaRe to write temporary mount point data to disk. Point (ii) is enabled by setting the \texttt{TMPDIR} environment variable to a disk mount, in the Apache Zeppelin configuration. Even if this is not optimal in terms of performance, it is necessary as the full partition size exceeds the tmpfs capacity in our worker nodes. Point (i) is implemented by using the \texttt{repartitionBy} primitive, on line 11. In particular, we specify a \texttt{keyBy} function that parses and returns a the chromosome identifier (on line 12), and a number of partitions that is equal to the number of worker nodes (on line 13).

The \texttt{map} primitive (on line 14) uses the chromosome-wise partitioning to perform the SNP calling, with GATK. Since the data is in SAM format, we set the input mount point as text file (line 15). However, since we are going to zip the results before aggregating the SNPs (line 32), we set the output mount point as a binary files directory (\texttt{"/out"}, on line 16). On line 17, we set the same Docker image that we used for the initial mapping step and, on line 18, we specify a command that: (i) prepends the necessary SAM header to the input data (which is available inside the container under \texttt{/ref/human\_g1k\_v37.dic}, on line 19), (ii) coverts the SAM input in BAM format (line 23), (iii) builds an index for the BAM format (line 26) and (iv) runs the multithreaded SNP calling using GATK, producing a Variant Call Format (VCF) file \cite{danecek2011variant} (line 28). A detailed description of the options, used for each command, can be found in the GATK documentation \cite{gatkdoc}.


Finally, to aggregate the SNPs to a single zipped file, we use the \texttt{reduce} primitive. In this case we use binary file mount points (lines 35 and 36) and a publicly available image containing the VCFtools software \cite{danecek2011variant} (line 37). On line 39, the specified command uses \texttt{vcf-concat} to merge all of the VCF files in the input mount point, and then it zips and writes them to the output mount point (line 40). Since MaRe applies the reduce command iteratively, intermediate partitions will contain multiple files. Therefore, to avoid file-name clashes, we include a random identifier in the command output (\texttt{\$RANDOM} at line 40). 

\begin{figure*}
    \centering
    \includegraphics[width=0.5\textwidth]{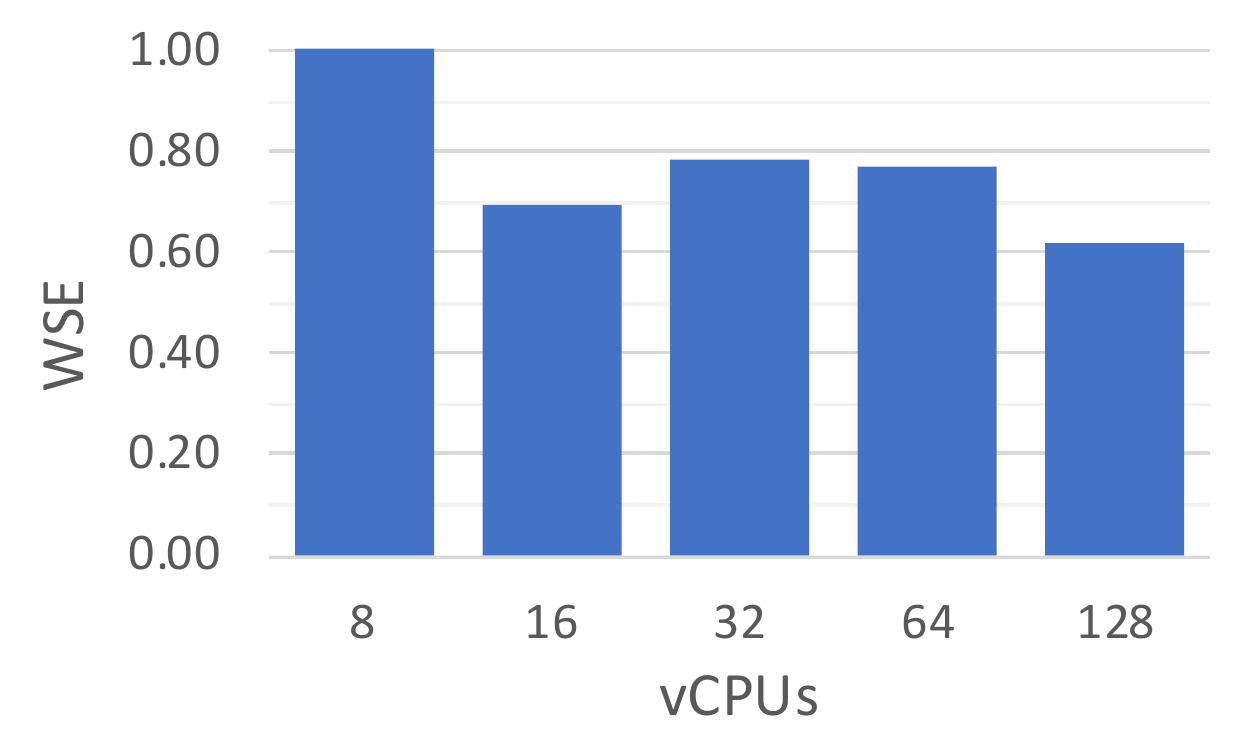}
    \caption{WSE for the SNP calling implemented in MaRe (listing \ref{lst:snp}). The results are produced by using a full individual dataset from the 1000 genomes project as input. Please notice that the \textit{vCPUs} axis is in base two logarithmic scale.}
    \label{fig:snp_wse}
\end{figure*}
\begin{figure*}
    \centering
    \includegraphics[width=0.5\textwidth]{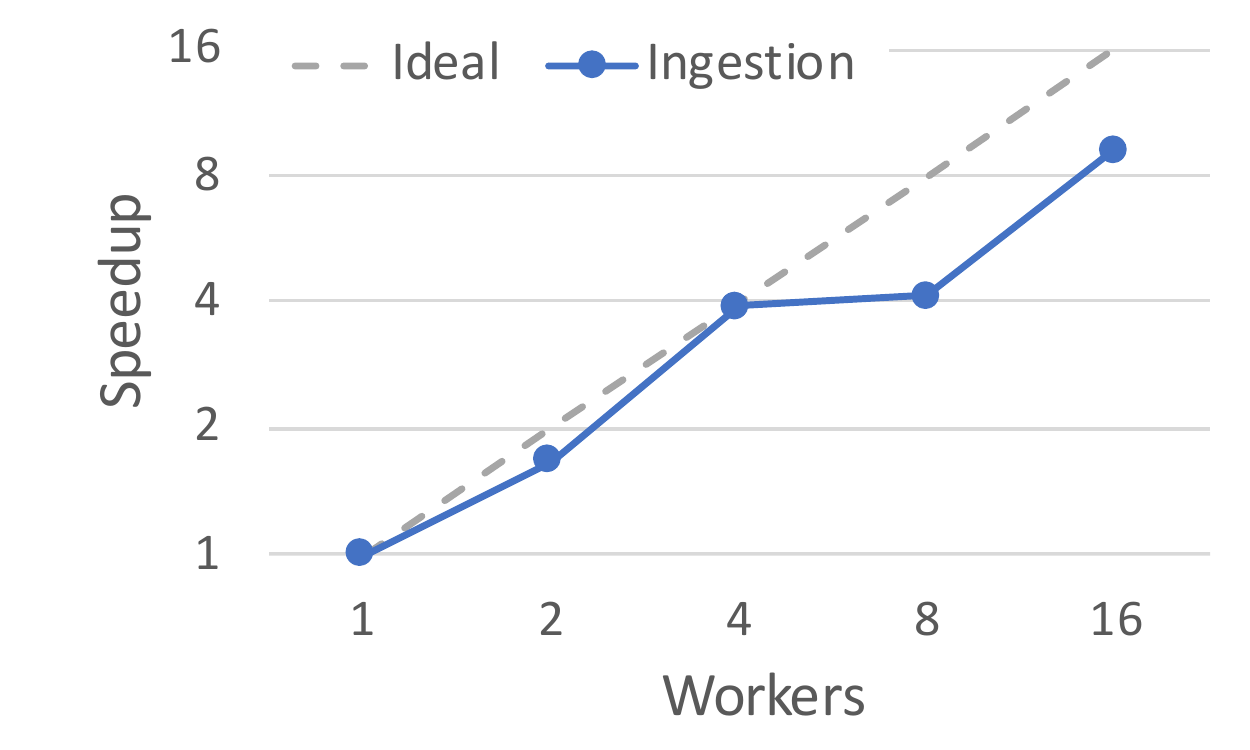}
    \caption{Ingestion speedup for one full individual dataset from the 1000 genomes project. Please notice that both axis are in base two logarithmic scale.}
    \label{fig:snp_speedup}
\end{figure*}

We benchmarked the analysis in listing \ref{lst:snp} against the full individual reads dataset \texttt{HG02666} ($\sim$30GB compressed FASTQ files), from the 1000 Genomes Project (1KGP) \cite{10002015global}. Since the tools run multiple threads, we configured the "spark.task.cpus" property to 8 in order to ensure proper resource allocation in our cluster setup. Amazon S3 hosts the full 1KGP dataset (\url{s3.amazonaws.com/1000genomes}) and represents a common ingestion source for this use case. The full benchmark runs in $\sim$1.8 hours when using 128 vCPUs, including data ingestion from S3. When studying the WSE for this application we do not consider the ingestion time, as obviously S3 does not host random samples of the dataset. In fact, when performing the runs to compute the WSE, we downsampled the data at run time. Under this assumption the input size of the ingestion phase is static. Therefore, instead of evaluating the ingestion in terms of WSE, we show how the speed increased when adding worker nodes. Figure \ref{fig:snp_wse} shows the WSE of the analysis (excluding ingestion) and figure \ref{fig:snp_speedup} shows the speedup of the ingestion phase. The speedup is computed as the ingestion time for N workers divided by the ingestion time for 1 worker. The WSE oscillates between 0.70 and 0.80 up to 64 vCPUs, and it decreases to $\sim$0.6 at 128 vCPUs. Even if this does not show optimal performance, as in the VS use case, it still indicates good scalability. The ingestion speedup is close to ideal for up to 4 workers, and it levels off slightly from 8 to 16 workers.

\subsection{Discussion and conclusions}
Big Data applications are getting increasing momentum in life science. Data is nowadays stored and processed in distributed systems, often in a geographically dispersed manner. 
This introduces a layer of complexity that MapReduce frameworks, such as Apache Spark, excel at handling \cite{khanam2015map}. Container engines, and in particular Docker, are also becoming an essential part of bioinformatics pipelines as they improve delivery, interoperability and reproducibility of scientific analyses. 
By enabling application containers in MapReduce, MaRe constitutes an important advancement in the scientific data-processing software ecosystem. 
When compared to current best practices in bioinformatics, relying solely on using workflow systems to orchestrate data pipelines, MaRe has the advantage of providing locality-aware scheduling, transparent ingestion from heterogeneous storage systems and interactivity. As data becomes larger and more globally distributed, we envision scientists to instantiate MaRe close to the data, and perform interactive analyses via cloud-oriented resources. In addition to the interactive mode, MaRe also support batch-oriented processing. This is important as it enables integration with existing bioinformatics pipelines. In practical terms, a packaged MaRe application can be launched by a workflow engine to enable data-intensive phases in a pipeline, and submitted to any of the resource managers supported by the Apache Spark community (including HPC systems \cite{chaimov2016scaling}).

In the evaluation section we have shown how researchers can easily implement two widely-used applications in life science, using MaRe. Both analyses can be coded in less than 50 lines of code, and they are seamlessly parallelized. 
The results show near optimal scalability for the VS application, with HDFS performing slightly better than Swift. The slight performance improvement is because of HDFS co-location with the worker nodes, allowing for less network communication. 

Scalability in the SNP calling analysis is reasonably good but far from optimal. The reason for this is that before running the haplotype caller, a reasonable amount of data needs to be shuffled across the nodes as GATK needs to see all of the data for a single chromosome at once in order to function properly, thus causing a large amount of data to be materialized on disk. Such overhead can be partly mitigated by enabling data streams via standard input and output between MaRe and containers, which constitutes an area for future improvement. Pure Apache Spark implementations of SNP calling such as ADAM~\cite{massie2013adam} show better scalability than MaRe. It is however important to point out that while ADAM is application specific, MaRe applies to a variety of use cases in bioinformatics and it stands out by enabling distributed SNP calling in less than 50 lines of code. To this extent, ADAM is the product of a large collaboration, counting thousands of lines of code. As such effort is not always sustainable, we point out that MaRe provides a reasonably good way of implementing general-purpose data-intensive pipelines with considerably smaller effort. Finally we highlight that, as opposite to ADAM, MaRe provides interoperability with any bioinformatics tool through the adoption of application containers (including ADAM itself).

In conclusion, MaRe provides a MapReduce-oriented model to enable container-based bioinformatics analyses at scale. The project is available on GitHub \cite{mare} under an open source license, along with all of the code to reproduce the analyses in the evaluation section \cite{benchmarks}.

\section{Methods}

\subsection{Apache Spark}
Apache Spark is an open source cluster-computing framework, for the analysis of large-scale dataset \cite{zaharia2010spark}. The project originally started with the aim of overcoming lack of in-memory processing in traditional MapReduce frameworks. Today, Apache Spark has evolved in a unified analytics engine, encompassing high-level APIs for machine learning, streaming, graph processing and SQL, and it has become the largest open source project in Big Data analytics with over 1000 contributors and over 1000 adopting organizations \cite{zaharia2016apache}.

\subsubsection{Clustering model}
The Apache Spark clustering model includes: a driver program, one or more worker nodes and a cluster manager. The driver program is written by the user and controls the flow of the programmed analysis. For interactive analysis the driver program can run in notebooks environments such as Jupyter \cite{kluyver2016jupyter} and Apache Zeppelin \cite{cheng2018building}. Worker nodes communicate with the driver program, thus executing the distributed analysis as defined by the user. Finally, a cluster manager handles resources in the cluster, allowing for the executing processes to acquire them in the worker nodes. Apache Spark is cluster-manager agnostic and it can run in stand alone settings, as well as on some popular platforms (e.g., Kubernetes \cite{kubernetes}, Mesos \cite{hindman2011mesos} and Hadoop YARN \cite{yarn}).

\subsubsection{Resilient Distributed Datasets}
Resilient Distributed Datasets (RDDs) \cite{zaharia2012resilient} are central in the Apache Spark programming model. RDDs are an abstraction of a dataset that is partitioned across the worker nodes. Hence, partitions can be operated in parallel in a scalable and fault-tolerant manner, and possibly cached in memory for recurrent access. As a unified processing engine, Apache Spark offers support for ingesting RDDs from numerous big-data-oriented storage systems. RDDs can be operated through: Scala \cite{odersky2004overview}, Python \cite{python}, Java \cite{java} and R \cite{ihaka1996r} APIs. Such APIs expose RDDs as object collections, and they offer high-level methods to transform the datasets.

The \textit{mapPartition} and the \textit{repartition} methods, from the RDD API, are useful to understand the MaRe implementation. The \textit{mapPartition} method is inspired by functional programming languages. It takes as an argument a lambda expression that codes a data transformation, and it applies it to each partition, returning a new RDD. The \textit{repartition} method, as the name suggests, changes the way the dataset records are partitioned across the worker nodes. It can be used to increase and decrease the number of partitions, thus affecting the level of parallelism, and it can also sort records in partitions, according to custom logics. In this case, an additional RDD method, namely \textit{keyBy}, needs to be used to compute a key for each RDD record. Similarly to \textit{mapPartition}, \textit{keyBy} applies a user-provided lambda expression to compute the record keys. Such keys are then used by \textit{repartition} in conjunction with an extension of the \textit{Partitioner} class \cite{hashpartitioner} to assign records to partitions. For instance, when using \textit{HashPartitioner} \cite{partitioner} records with same key always end up in the same RDD partition. 
 
\subsubsection{Stages and data locality}
RDD methods are lazily applied to the underlying dataset. This means that until something needs to be written to a storage system, or returned to the driver program, nothing is computed. In this way, Apache Spark can build a direct acyclic graph and thus optimize the physical execution plan. A physical execution plan is composed of processing tasks that are organized in stages. Typically, inside each stage the physical execution plan preserves data locality, while between stages a data shuffle occurs. In particular, a sequence of {mapPartition} methods generate a single stage, giving place to almost no communication in the physical execution plan. In contrast, each time \textit{repartition} is applied to an RDD, a new stage is generated (and data shuffling occurs).

\subsection{Docker}
Docker has emerged as the de-facto standard application container engine \cite{shimel2016docker}. Like Virtual Machines (VMs), application containers enable the encapsulation of software components so that any compliant computer system can execute them with no additional dependencies \cite{oci2016}. The advantage of Docker and similar container engines over virtualization consists of eliminating the need of running an Operative System (OS) for each isolated environment. As opposite to hypervisors, container engines leverage on kernel namespaces to isolate software environments, and thus run containers straight on the host OS. This makes application containers considerably lighter than VMs, enabling a more granular compartmentalization of software components.

\subsubsection{Software Delivery}
By enabling the encapsulation of entire software stacks, container engines have the potential of considerably simplify application delivery. Engines such as LXC \cite{lxc} and Jails \cite{kamp2000jails} have been available for almost two decades. Nevertheless, when compared to Docker these systems are poor in terms of software delivery functionalities. This is the reason why software containers popularity exploded only when Docker emerged.

Docker containers can be defined using a text specification language. Using such language, users compose a \textit{Dockerfile} which is parsed by Docker, and then complied into a Docker image. Docker images can then be released to public or private registries, becoming immediately available over the Internet. Therefore, by running the Docker engine, the end users can conveniently start the released containers locally.

\subsubsection{Volumes}
When using Docker containers for data processing, volumes play an important role. Indeed, there is a need for a mechanism to pass the input data to the containers, and to retrieve the processed output from the isolated environment. Docker volumes allow for defining shared file spaces between containers and the host OS. Such volumes can be easily created when starting containers, by specifying a mapping between host OS file, or directories, and container mount points. Inside the containers these shared objects simply appear as regular files, or directories, under the specified mount point.

\section{Availability of supporting source code and requirements}
Project name: MaRe

\noindent Project home page: \url{https://github.com/mcapuccini/MaRe}

\noindent Operating system(s): Platform independent

\noindent Programming language: Scala

\noindent Other requirements: Apache Spark and Docker

\noindent License: Apache License 2.0

\section{Availability of supporting data}
The data set supporting the VS evaluation in this article is available in the ZINC database \cite{irwin2012zinc}. The data set supporting the SNP evaluation is available on Amazon S3 (\url{s3.amazonaws.com/1000genomes}).

\section{Declarations}

\subsection{List of abbreviations}
1KGP: one thousand genome project; API: application programming interface; CSC: information technology center for science; HDFS: hadoop distributed file system; HPC: high-performance computing; OS: operative system; RDD: resilient distributed dataset; SDF: structure-data file; SNP: single nucleotide polymorphism; VCF: variant call format; VM: virtual machine; VS: virtual screening; WSE: weak scaling efficiency; vCPU: virtual central processing unit.

\subsection{Ethics approval and consent to participate}
All of the 1KGP data is consented for analysis, publication and distribution. Ethics and consents are extensively explained in the 1KGP publications \cite{10002015global}.

\subsection{Competing interests}
The authors declare that they have no competing interests.

\subsection{Funding}
This research was supported by The European Commission’s Horizon 2020
programme under grant agreement number 654241 (PhenoMeNal).

\subsection{Author's contributions}
MC and OS conceived the project. MC designed and implemented MaRe. MC and MD carried out the evaluation experiments. MD provided expertise in genomics. ST provided expertise in cloud computing. All authors read and approved the final manuscript.

\subsection{Acknowledgment}
We kindly acknowledge contributions to cloud resources by CSC (\url{https://www.csc.fi}), the Nordic e-Infrastructure Collaboration (\url{https://neic.no}) and the SNIC Science Cloud \cite{toor2017snic}.

\bibliography{references}
\bibliographystyle{ieeetr}

\clearpage

\end{document}